\documentclass[a4paper, oneside, twocolumn, notitlepage, 10pt]{extarticle_ecoc}
\usepackage{ecoc}

\begin{document}
\selectlanguage{english}    % Standard Language

%-------------------------------------------------- Title -----------------------------------------------------%

\title{Gain Profile Characterization and Modelling\\for an Accurate EDFA Abstraction and Control}%

%------------------------------------------------- Authors-----------------------------------------------------%

\author{
    Giacomo Borraccini\textsuperscript{(1)}, 
    Vittorio Gatto\textsuperscript{(1)}, 
    Andrea D'Amico\textsuperscript{(1)},
    Stefano Straullu\textsuperscript{(2)},
    Francesco Aquilino\textsuperscript{(2)}, \\
    Stefano Piciaccia\textsuperscript{(3)},
    Alberto Tanzi\textsuperscript{(3)},
    Gabriele Galimberti\textsuperscript{(4)} and
    Vittorio Curri \textsuperscript{(1)}
}

\maketitle                  % Create title and author

%------------------------------------------ Description of Authors ----------------------------------------------%

\begin{strip}
 \begin{author_descr}

   \textsuperscript{(1)}Politecnico di Torino, Turin, Italy,
   \textcolor{blue}{\uline{giacomo.borraccini@polito.it}}; \textsuperscript{(2)}LINKS Foundation, Turin, Italy; \textsuperscript{(3)}Cisco Photonics, Vimercate, Italy; \textsuperscript{(4)}Internet Engineering Task Force (IETF)

 \end{author_descr}
\end{strip}

\setstretch{1.1}
%-------------------------------------------------- Footnote -------------------------------------------------------%
\renewcommand\footnotemark{}
\renewcommand\footnoterule{}
%\let\thefootnote\relax\footnotetext{text}

%-------------------------------------------------- Abstract ---------------------------------------------------------%

\begin{strip}
  \begin{ecoc_abstract}
    % NOTE: Don't use a blank line here but start abstract right away to avoid an extra line break
Relying on a two-measurement characterization phase, a gain profile model for dual-stage EDFAs is presented and validated in full spectral load condition.
It precisely reproduces the EDFA dynamics varying the target gain and tilts parameters as shown experimentally on two commercial items from different vendors. \textcopyright2023 The Author(s)
\vspace{-2mm}
  \end{ecoc_abstract}
\end{strip}

%-------------------------------------------------- Introduction Section -------------------------------------------------------%

\section{Introduction}
Maximizing the capacity of optical infrastructures is one of the main objectives for operators and service providers, aiming at minimizing costs at the same time~\cite{riccardi2018operator}.
For this purpose, quality-of-transmission estimation (QoT-E) represents a fundamental aspect for both the optical control and data planes, and predicting the expected behavior of the system with a reasonable margin~\cite{pointurier2017design}.
To this purpose, with an open networking perspective, the open source GNPy model \cite{curri2022gnpy} is an extensively validated vendor-agnostic solution.
As a result, optical amplifiers are key network elements that allow signal power levels to be restored at the cost of signal-to-noise ratio degradation.
The characterisation of optical amplifiers includes two main parameters that are critical in the search for the optimum operating point of the system: the gain profile, $g$, and the noise figure profile.
In general, an adequate description of the frequency-dependent physical layer parameters plays an important role in the QoT-E, especially when considering wideband transmission scenarios~\cite{damicoJLT2022}.
Focusing on Erbium-doped fiber amplifiers (EDFAs), an ideal gain profile can be expressed in decibels as:
\vspace{-0.1cm}
\begin{equation}\label{eq:e}
    g(f; G, T) = G + \frac{T}{B}(f -f_c)
\end{equation}
where $f$ is the optical frequency,  $G$ and $T$  are the target gain and tilt parameters set in the amplifier, $f_c$ is the rotation pivot point of the tilt, and $B$ is the operative EDFA bandwidth. 

Accurate application of the tilt is important to compensate for stimulated Raman scattering (SRS) due to the propagation through the fiber span~\cite{Rottwitt:03}.
Eq.~\ref{eq:e} completely neglects the additional gain profile ripple due to the EDFA manufacturing and the physical behaviour.
The design of the EDFA gain profile can be achieved through the evaluation of the dynamic gain tilt (DGT) parameter, evaluated as the normalized difference, with respect to a specific frequency of two different gain profiles~\cite{article:dgt}.  
Starting from the DGT definition and from the explicit interpretation of the gain profile, it is possible to define a constant parameter regardless of the specific configuration of tilt and gain parameters that is set in the amplifier.
Different machine learning (ML) strategies have already tackled the problem of accurately modelling the gain ripple profile~\cite{zhu2018machine, ionescu2019machine, mahajan2020modeling, zhu2020hybrid}, requiring relatively large datasets.
ML solutions can also compensate for gain profile fluctuations induced by a varying spectral load, as shown in~\cite{d2020using}.

In this work, a semi-analytical EDFA model is presented and validated, focusing on the accurate reproduction of the gain profile, including the gain ripple, in a full spectral load transmission scenario.
Extending Eq.~\ref{eq:e}, the proposed methodology provides a low computational cost procedure for the gain ripple profile evaluation. 
The implemented abstraction is composed of two different steps: first, the EDFA gain profile is characterized considering only two gain/tilt settings in full spectral load condition, that is the most interesting solution by on case exploiting ASE shaped channels.
Then, the semi-analytical model enables an accurate estimation of the gain profile for any pair of gain/tilt settings. 
This implementation is very promising for its precision and simplicity, fundamental features for efficient and agile QoT-E such as the one within GNPy~\cite{curri2022gnpy}.
\section{Methodology}
Without any loss of generality, an accurate EDFA gain profile modeling is described as follows:
\begin{equation}\label{eq:eq1}
    g(f; G, T) = G + \frac{T}{B}(f -f_c) + r_T(f)
\end{equation}
where the term $r_T(f)$ refers to the gain ripple profile, that can be defined as the deviation between the real and ideal amplifier gain profiles~\cite{borraccini2023iterative}.

In general, all the gain ripple profiles, $r_T(f)$, generated in a different tilt condition are dependent on the set target tilt parameter, $T$.
In particular, the gain ripple profile of a real EDFA evaluated when the tilt parameter is set to 0~dB is a manufacturing footprint of the EDFA design procedure and, indeed, does not depend on $T$:
\begin{equation}\label{eq:eq2}
    g(f; G, T=0) = G + r_0(f)\,.
\end{equation}
Bearing in mind the DGT-based EDFA design that describes the physical implementation of the tilt, it is possible to define a tilt-independent parameter, $K(f)$, defined as:
\begin{equation}\label{eq:eq3}
K(f) = \frac{r_0(f) - r_T(f)}{T}\,,
\end{equation}
which is a feature of the EDFA figure of merit, and $r_T(f)$ is derived from the gain profile measurement setting the tilt parameter at $T$. 
From Eq.~\ref{eq:eq3}, it is possible to evaluate a generic tilt-dependent gain ripple profile in function of the $r_0(f)$ and $K(f)$:
\begin{equation} \label{eq:eq6}
    r_T(f) = r_0(f) - T\,K(f)
\end{equation}

Using the Eq.~\ref{eq:eq6}, it is possible to obtain an accurate evaluation of a generic gain profile for a specific EDFA:

\begin{equation}\label{eq:eq4}
g(f; G, T) = G + \frac{T}{B}(f-f_c) + r_0(f) - T\,K(f)
\end{equation}

\begin{figure}[!b]
   \centering
    \includegraphics[width=0.8\linewidth]{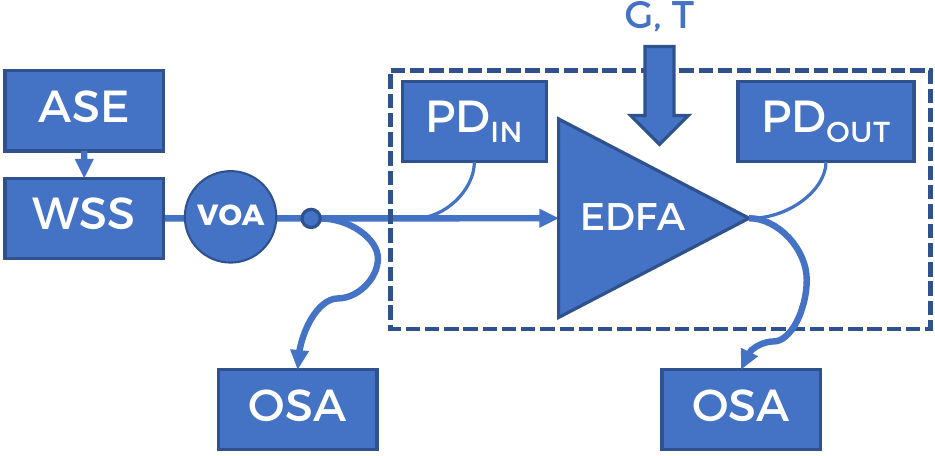}
    \caption{Sketch of the experimental setup used to characterize a single EDFA item.}
    \label{fig:characterization_setup}
\end{figure}

\begin{figure*}[!t]
   \centering
        \includegraphics[width=1\linewidth]{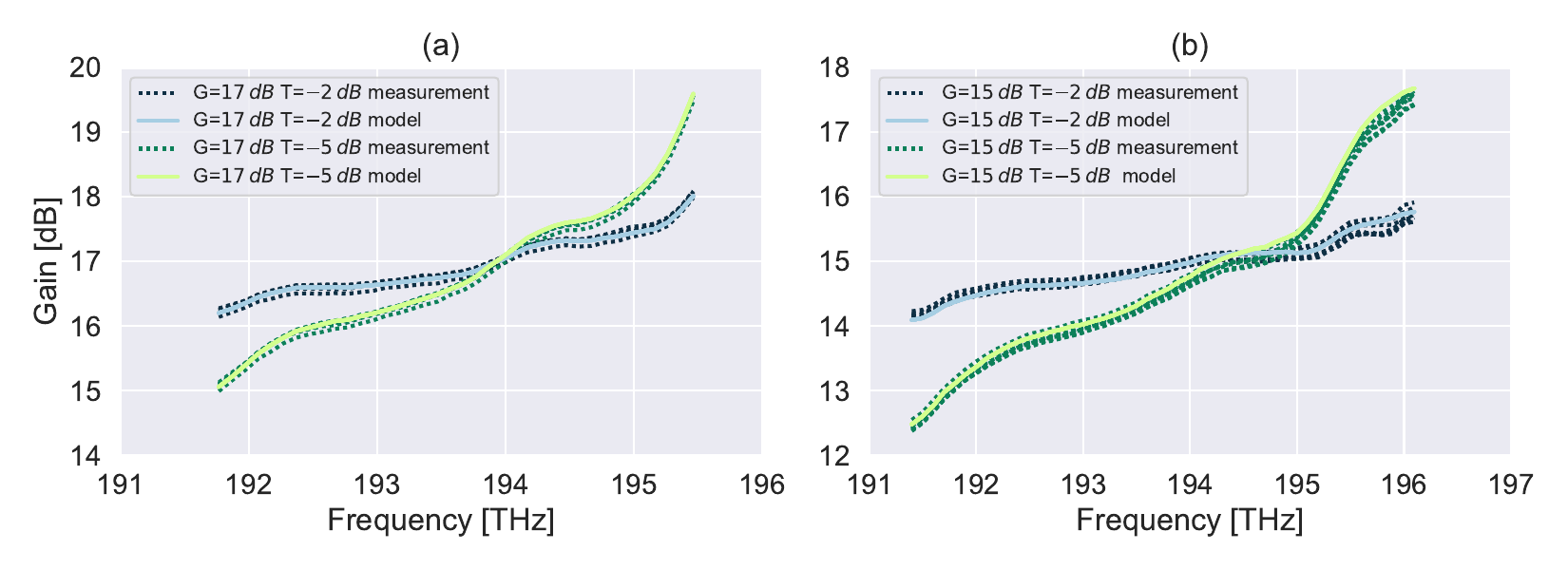}
    \caption{Gain profile comparison between the experimental measurements and the presented EDFA model for two different configurations: (a) EDFA~1; (b) EDFA~2.}
    \label{fig:edfa_profie}
\end{figure*}

 \vspace{-2mm}
\section{Measurements}

\begin{figure*}[!t]
   \centering
    \includegraphics[width=1\linewidth]{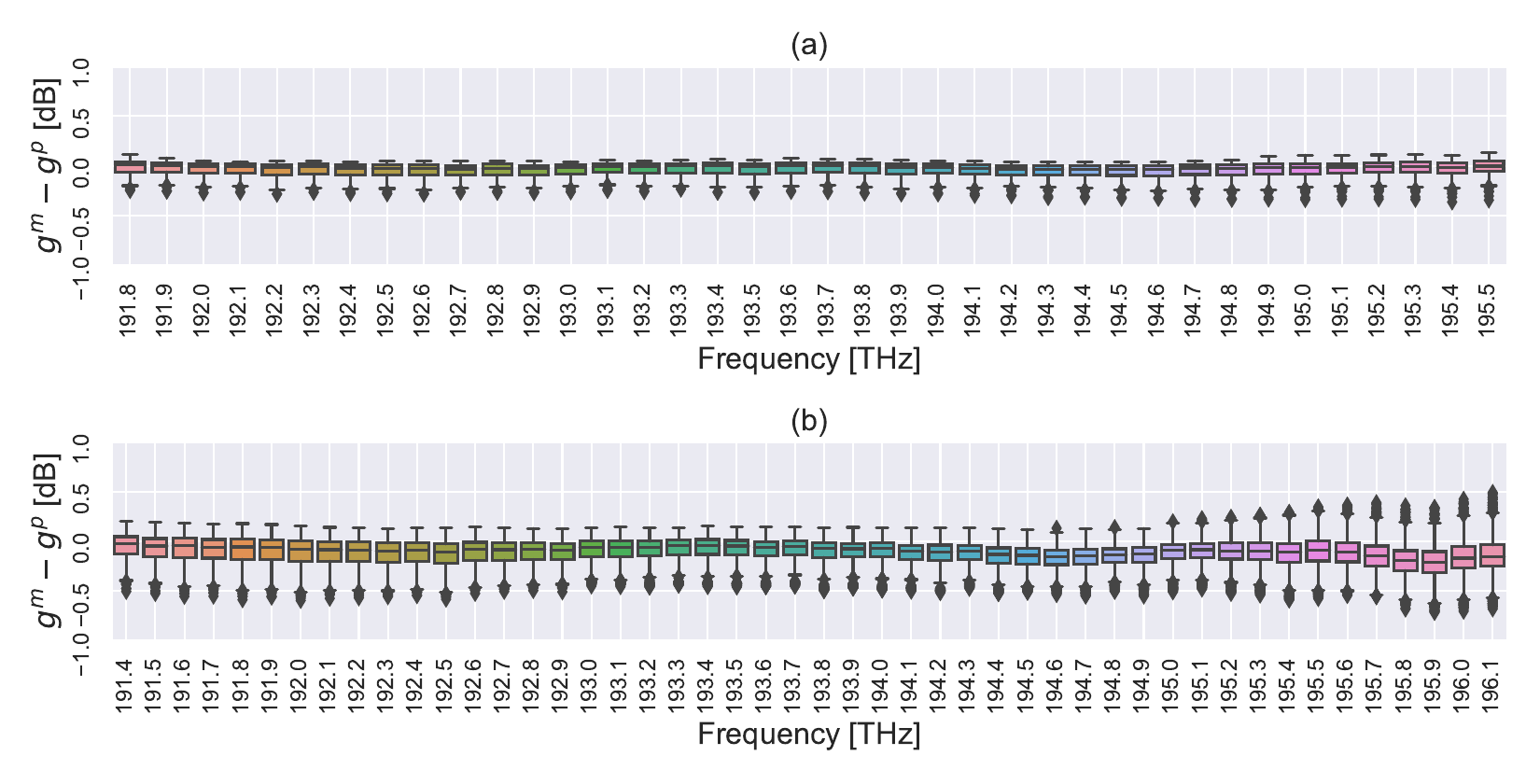}
    \caption{Box plot resuming the error estimation between the all the performed experimental measurements, $g^m$, and the corresponding predictions using the presented model, $g^p$: (a) EDFA~1; (b) EDFA~2.}
    \label{fig:error_comparison}
\end{figure*}

To validate the model, a full experimental characterization of two set of two distinct EDFA families (referred as EDFA~1 and EDFA~2 in the following) is performed.
For each item, an experimental dataset consisting of different gain profiles for a given combinations of total input power, target gain and tilt parameters, have been collected.
The configuration setup for the acquisition of the dataset is depicted in Fig.~\ref{fig:characterization_setup}.
A commercial wavelength selective switch (WSS) is programmed in order to shape the ASE noise generating a wavelength division multiplexed (WDM) comb of either 40 or 48 channels, 100~GHz spaced, each modulated at 32~GBaud; this way, two different spectral loads are provided at the EDFAs' input, according to each device specifications: 4~THz for EDFA~1, 4.8~THz for EDFA~2, respectively.
Each amplifier can be controlled via secure shell protocol (SSH), which has been exploited to set the gain and the tilt parameters, whereas its optical input power is changed acting on the variable optical attenuator (VOA) placed in front of it. 
The optical spectrum at both the EDFA's input and output is captured using an optical spectrum analyzer (OSA). 
The dataset for EDFA~1 has been collected changing the input power from -10~dBm to +6~dBm, with steps of 2~dB, and setting EDFA's gain and tilt values in the range [14:20]~dB and [-5:4]~dB, respectively, each with steps of 1~dB.
Regarding EDFA~2, the gain is extend to a range of to [12:27]~dB with a 1~dB step, consequently reducing the upper limit of the input power range to +4~dBm.

The presented model has been derived from both the EDFA items under investigation.
By fixing a target gain parameter value, the gain ripple imposing tilt equal to 0~dB, $r_0(f)$, has been derived simply by measuring the corresponding gain profile and removing the average gain.
Since the target tilt parameter is generally defined in wavelength, the $K(f)$ has been evaluated with the gain profile with tilt 0~dB and with the gain profile that has the larger negative target tilt parameter value. 
The choice of the second gain profile is arbitrary, but the experimental evidence has shown that a larger tilt results in a smaller relative measurement error.
Furthermore, considering the non-zero target tilt parameter, the tilt pivot point, $f_c$, is evaluated as the linear regression of the measured gain profile assumes the value of the target gain parameter.
The EDFA operating band, $B$, is estimated by considering the same linear regression so that the value of the set target tilt parameter agrees with the estimate.

\vspace{-2mm}
\section{Results}

The Eq.~\ref{eq:eq4} has been used to evaluate the gain profile with a specific gain target and a specific tilt target. 
In Fig.~\ref{fig:edfa_profie}, it is possible to observe the measured gain profile and the evaluated gain profile for both the considered EDFAs setting the different configurations of gain and parameter values. 
A single gain profile is provided using the proposed model for all the values of total input powers as it is independent with respect to the latter.
It is worth noting that the measured profiles are affected by measurement post-processing uncertainty, as all the gain profiles have been obtained as a differential calculation of the input/output power profiles, measured with the OSA, and re-scaled to the EDFA input/output total powers, which are affected by an inaccuracy of $\pm$0.1~dB.

Fig.~\ref{fig:error_comparison} provides a larger view of the model effectiveness, showing the estimated errors comparing the measured gain profiles and the gain profiles evaluated by the model for all the possible combinations of total input power, gain and tilt values.
In both cases, the error is centered in 0~dB, demonstrating an unbiased estimation, showing a worst case value of 0.1~dB for the EDFA~1, and roughly 0.3~dB for the EDFA~2, without considering the outliers.
Furthermore, it can be observe that the model error is slightly higher in the high-frequency amplification region.
This can be explained with the well-known spectral hole burning phenomenon \cite{bolshtyansky2003spectral}.

\section{Conclusions}

An accurate and simple gain profile modelling for dual-stage EDFAs working in full spectral load condition is proposed and experimentally verified on two classes of devices coming from different vendors.
In particular, by means of the characterization of two device-specific invariant profiles, $K(f)$ and $r_0(f)$, which can be evaluated with only two gain profile samples, all the gain profiles for all the combination of input power, gain and tilt values can be obtained with a semi-analytical expression, Eq.~\ref{eq:eq6}. 
As the characterization procedure does not include any complexity, it can be performed both by the vendors and final users, and the device-specific characterization can be easily stored or shared.

%-------------------------------------------------- Acknowledgements Section -------------------------------------------------------%
\clearpage

%-------------------------------------------------- Bibliography Section -------------------------------------------------------%

\printbibliography

\vspace{-4mm}

%%%%%%%%%%%%%%%%%%%%%%%%%%%%%%%%%%%%%%%%%%%%%
%---------------------------------------------- End of Document -----------------------------------------------%
\end{document}